\documentclass[10pt,conference]{IEEEtran}
\usepackage[a-1b]{pdfx}
\usepackage{booktabs} 
\usepackage{sty}         
\usepackage{shortcuts}
\usepackage{multirow}
\usepackage{color, colortbl}
\usepackage{multicol}
\usepackage{pifont}
\usepackage{tablefootnote}
\usepackage{wasysym}
\usepackage{graphicx}

\newcommand{\approach}{AQE\xspace}%
\newcommand{\fullapproach}{Automated Query Evaluation\xspace}%

\definecolor{realblue}{rgb}{0,0,1}

\hypersetup{
  colorlinks=true,
    linkcolor = realblue,
    anchorcolor = realblue,
    citecolor = realblue,
    filecolor = realblue,
    urlcolor = realblue
}

\usepackage{cite}
\usepackage{shortcuts}
\usepackage{listings}
\usepackage{csquotes}
\usepackage{listings-golang}
\usepackage{xcolor}
\usepackage{fancyvrb}
\usepackage[T1]{fontenc}

\usepackage[most]{tcolorbox}
\tcbuselibrary{skins,breakable,listings}

\newcommand{\si}[1]{\small\texttt{#1}\normalsize}

\newenvironment{blockquoted}{%
  \par%
  \medskip
  \leftskip=1em\rightskip=0em%
  \noindent\ignorespaces}{%
  \par\medskip}

\newcommand\vtextvisiblespace[1][.5em]{%
  \makebox[#1]{%
    \kern.07em
    \vrule height.3ex
    \hrulefill
    \vrule height.3ex
    \kern.07em
  }
}

\tcbset{
    boxrule=0pt,
    left=0pt,
    right=0pt,
    top=0pt,
    bottom=0pt,
    }

\definecolor{dkgreen}{rgb}{0,0.5,0}
\definecolor{dkred}{rgb}{0.5,0,0}
\definecolor{gray}{rgb}{0.5,0.5,0.5}
\definecolor{vlgray}{gray}{0.95}
\definecolor{lgray}{gray}{0.7}
\definecolor{bluehighlight}{HTML}{46adb7}
\definecolor{orangehighlight}{HTML}{e3a24d}
\definecolor{redhighlight}{HTML}{ffeef0}
\definecolor{greenhighlight}{HTML}{e6ffed}
\definecolor{grayhighlight}{HTML}{6a737d}
\definecolor{blu}{HTML}{035cc5}
\definecolor{LLGray}{gray}{0.93}

\newcommand{\codepurple}[1]{\texttt{\small\color{purple}#1}}
\newcommand{\codeblue}[1]{\texttt{\small\color{blue}#1}}
\newcommand{\codeactuallyblue}[1]{\texttt{\small\color{blue}#1}}
\newcommand{\codeblack}[1]{\texttt{\small#1}}
\newcommand{\codegray}[1]{\texttt{\small\color{gray}#1}}

\newcommand{\ts}[1]{\texttt{\small {#1}}}

\newcommand{\mc}[3]{\multicolumn{#1}{#2}{#3}}

\newcommand\lt[1]{{\lstinline!#1!}}
\renewcommand\t[1]{{\lstinline!#1!}}

\tcbuselibrary{skins,breakable,listings}

\lstdefinestyle{comby}{
  basicstyle=\footnotesize\ttfamily,
  numbers=none,
  escapeinside={@?}{?@},
  classoffset=0,
  keywordstyle=\color{blue},
  classoffset=1,
  keywordstyle=\color{orange},
  morekeywords={val,let,fun,match,with,rec,raise,return,type},
  classoffset=2,
  keywordstyle=\color{purple},
  morekeywords={list,string,unit,error},
  classoffset=3,
  morecomment=[s]{(**}{*)},
  moredelim=[is][\color{blue}]{@X}{X@},
  moredelim=[is][\color{purple}]{@Y}{Y@},
  escapeinside={(?}{?)},
}

\lstdefinestyle{gostyle}{
  basicstyle=\ttfamily\bfseries\scriptsize\linespread{1.1},
  frameround=tttt,
  frame=single,
  keywordstyle=\color{blue},
  commentstyle=\color{dkred},
  stringstyle=\color{dkgreen},
  keepspaces=true,              
  numbers=left,
  breaklines=true,
  otherkeywords={::=},
  numberstyle=\ttfamily\footnotesize\color{gray},
  stepnumber=1,
  numbersep=8pt,
  backgroundcolor=\color{white},
  tabsize=4,
  showspaces=false,
  showstringspaces=false,
  xleftmargin=.18in,
  captionpos=b,
  escapeinside={(?}{?)},
}

\lstdefinestyle{highlightdelims}{
  basicstyle=\footnotesize\ttfamily,
  numbers=none,
  escapeinside={@?}{?@},
  classoffset=0,
  keywordstyle=\color{blue},
  otherkeywords={->,>>,=,<|>,::,[]},
  classoffset=1,
  keywordstyle=\color{orange},
  morekeywords={val,let,fun,match,with,rec,raise,return,type},
  classoffset=2,
  keywordstyle=\color{purple},
  morekeywords={list,string,unit,error},
  classoffset=3,
  morecomment=[s]{(**}{*)},
  moredelim=[is][\color{red}]{@X}{X@},
}

\lstdefinestyle{cstyle}{
language=c,
basicstyle=\ttfamily\bfseries\scriptsize,
  morekeywords={virtualinvoke},
  keywordstyle=\color{blue},
  ndkeywordstyle=\color{red},
  commentstyle=\color{dkred},
  stringstyle=\color{dkgreen},
  keepspaces=true,              
  numbers=left,
  breaklines=true,
  numberstyle=\ttfamily\footnotesize\color{gray},
  stepnumber=1,
  numbersep=10pt,
  backgroundcolor=\color{white},
  tabsize=4,
  showspaces=false,
  showstringspaces=false,
  xleftmargin=.23in,
  captionpos=b,
  escapeinside={(?}{?)},
  escapeinside={$}{$}
}

\lstdefinestyle{spacestyle}{
basicstyle=\ttfamily\bfseries\scriptsize,
  morekeywords={virtualinvoke},
  keywordstyle=\color{blue},
  ndkeywordstyle=\color{red},
  commentstyle=\color{dkred},
  stringstyle=\color{dkgreen},
  keepspaces=true,
  numbers=left,
  breaklines=true,
  numberstyle=\ttfamily\footnotesize\color{gray},
  stepnumber=1,
  numbersep=10pt,
  backgroundcolor=\color{white},
  tabsize=4,
  showspaces=true,
  showstringspaces=true,
  xleftmargin=.23in,
  captionpos=b,
}

\begin{document}

\title{You Don't Know Search: Helping Users Find Code by Automatically Evaluating Alternative Queries}

\author{\IEEEauthorblockN{Rijnard van Tonder}
\IEEEauthorblockA{\textit{Sourcegraph, Inc.} \\
San Francisco, USA \\
rijnard@sourcegraph.com}
}

\maketitle
\lstset{style=gostyle}

\lstdefinestyle{cstyle}{
language=c,
basicstyle=\ttfamily\bfseries\scriptsize,
  morekeywords={virtualinvoke},
  keywordstyle=\color{blue},
  ndkeywordstyle=\color{red},
  commentstyle=\color{dkred},
  keepspaces=true,              
  numbers=left,
  breaklines=true,
  numberstyle=\ttfamily\footnotesize\color{gray},
  stepnumber=1,
  numbersep=10pt,
  backgroundcolor=\color{white},
  tabsize=4,
  showspaces=false,
  showstringspaces=false,
  xleftmargin=.23in,
  captionpos=b,
  escapeinside={(?}{?)},
  frame=none,
}

\lstset{style=cstyle}

\definecolor{dkblue}{rgb}{0,0,.6}
\definecolor{dkyellow}{cmyk}{0,0,.8,.3}

\begin{abstract}
Tens of thousands of engineers use Sourcegraph day-to-day to search for code and
rely on it to make progress on software development tasks. We face a key
challenge in designing a query language that accommodates the needs of a broad
spectrum of users.
Our experience shows that users express different and often contradictory
preferences for how queries \emph{should} be interpreted. These preferences stem
from users with differing usage contexts, technical experience, and implicit
expectations from using prior tools. At the same time, designing a code search
query language poses unique challenges because it intersects traditional search
engines and full-fledged programming languages. For example, code search queries
adopt certain syntactic conventions in the interest of simplicity and terseness
but invariably risk encoding implicit semantics that are ambiguous at face-value
(a single space in a query could mean three or more semantically different
things depending on surrounding terms). Users often need to disambiguate intent
with additional syntax so that a query expresses what they actually want to
search. This need to disambiguate is one of the primary frustrations we've seen users experience with writing search queries in the
last three years. We share our observations that lead us to a fresh perspective
where code search behavior can straddle seemingly ambiguous queries. We develop
\fullapproach (\approach), a new technique that automatically generates and adaptively runs alternative query
interpretations in frustration-prone conditions.
We evaluate \approach
with an A/B test across more than 10,000 unique users on our publicly-available code
search instance. Our main result shows that relative to the control group, users are on average \textbf{22\%}
more likely to click on a search result at all on any given day when \approach is active.
We share our technique, learnings, and implementation that made it possible for a substantial
number of users to now see and click on results that they would not have seen otherwise.
\end{abstract}

\section{Introduction}
\label{sec:intro}

Sourcegraph\footnote{\href{https://about.sourcegraph.com}{about.sourcegraph.com}}
is a code search engine used by over 1 million engineers, and used at companies
like Dropbox, Cloudflare, Uber, Reddit, and many more. A free, open-source
instance of Sourcegraph also powers code search for millions of open source
repositories.\footnote{\href{https://sourcegraph.com/search}{sourcegraph.com/search}}
Developers use Sourcegraph to reference existing implementations, navigate
code, or find usage examples, in line with previous studies on how developers
use code search~\cite{howcodesearch}. Every user starts a code search by typing
a query into Sourcegraph's search bar. This single search bar encodes all the
expressive power for filtering repositories or files, searching with regular
expressions, or applying boolean operators (\codepurple{AND}, \codepurple{OR}) to search
terms. A user's familiarity with the query language influences whether they
can write a query that returns meaningful results. Previous studies show
developers frequently reformulate queries and run successive searches in a short
time span to home in on a desired result~\cite{howcodesearch}. At the same time, we've seen a broad spectrum
of users search code with Sourcegraph. First-time users may be unfamiliar with
query filters, or even general pattern matching with regular expressions. Even
familiar users will tweak queries to get results. 
User interaction with a query string across varying degrees of proficiency underscores a
single important design principle: help users find useful results quickly by (a)
surfacing the expressive capabilities of the language, and (b)
reducing the 
burden to modify or reformulate queries.
Beyond learning search syntax, Sourcegraph users regularly express
frustration via our feedback channels that queries aren't \emph{interpreted} the
way they expect.


In the past three years we've seen users express the gamut of differing personal
preferences for how Sourcegraph \emph{should} interpret certain
queries. At the surface level, designing a consistent and
predictable language for searching code seems deceptively simple. Our experience
reveals that query design presents one of the most contentious and difficult
challenges for a code search solution---not because it is difficult to design a
query language with consistent syntax and semantics, but because users regularly
express contradictory preferences on how queries should be interpreted.

\emph{To quote or not to quote? An example to illustrate complexity in interpreting code search queries.}
We often see Sourcegraph users get
tripped up by whether quotes in the search string are treated literally or not.
Consider two users who search for a quoted string like
\codeblue{"v1.3"}. User \#1 quotes \codeblue{v1.3} because
they want to search for that exact string (perhaps leaning on their experience
using Google search). User \#2 enters a quoted string because they want to find
that string \emph{including} quotes (it helps them narrow down results to a JSON
configuration file like \codeblack{"version":} \codeblue{"v1.3"}). Either
interpretation is reasonable, and users will expect or prefer certain behavior
depending on e.g., previous experiences, tools, or the task at hand when running
the search. Nevertheless both these users exist, and they are on opposite sides
of the ``interpretation camp''. Unfortunately, from a language design perspective,
there is no objectively better
way to interpret the presence of quotes in this scenario---either may be
preferred over the other. In practice, attempting to resolve this issue by allowing users to configure individual preferences for query behavior is highly problematic in a
collaborative context. For example, users could share syntactically identical
queries with others, who then see different or ``wrong'' results based on their
configuration (we elaborate on this complexity in Section \S\ref{sec:background}). Thus
the tool builder faces a crucial challenge: ambiguity must be resolved
so the language is consistent (we decide either quotes are literal or not),
and immediately once resolved, the language now grants a particular convenience
to one type of user, while inconveniencing the other. For example, supposing we decide
quotes are \emph{not} treated literally, User \#1 perceives no friction, but User \#2
sees unwanted results (maybe a comment containing the string
\codeblue{v1.3}). User \#2 now has to reformulate their query to
escape the quotes (e.g., \codeblue{\textbackslash"v1.3\textbackslash"} or
\codeblue{"\textbackslash"v1.3\textbackslash""}). Going the other way, where
quotes \emph{are} significant, User \#2 perceives no friction, but User \#1 sees
results including quotes, or worse, they may not see \emph{any} results because
there are no matches that include quotes (and they end up missing results for
the \codeblue{v1.3} pattern they were looking for). This time User \#1 will have
to modify their query.

Despite this complexity, a best-in-class code search tool must rise to the challenge to accommodate all
potential users: It must find ways to
reconcile contradictory expectations, educate users about capabilities, and
reduce query modification and friction. This is an acute challenge we've faced
and a demanding technical task facing every code search tool today. Recently we've
approached this problem with fresh insight, prompting a new solution to \emph{automatically
generate and run alternative interpretations}. The rest of this paper proceeds with the following outline and contributions:

\begin{blockquoted}
\textbf{Principles for code search design re: interpreting queries.} Section \ref{sec:background} shares
 background principles and scope behind building a code search engine (query
 syntax, expressive power, result ordering) that bear specifically on query
 interpretation.
\end{blockquoted}

\begin{blockquoted}
\textbf{Practical challenges in query interpretation.}
  Section \ref{sec:challenge} details our domain-specific
  challenges when developing query languages for an industrial-strength code search tool. We've condensed hundreds of pieces of user feedback to identify ambiguity in user expectations
  and how these relate to a consistent query syntax and semantics. We believe
  our findings are general and tool-agnostic, with relevance to design choices behavior found
  in all the major existing code search
  tools (e.g.,~\cite{google-chromium-search}, \cite{gh-beta-search}, \cite{opengrok-lucene}) and those that will appear in future.
\end{blockquoted}

\begin{blockquoted}
\textbf{A new approach for \fullapproach (\approach).}
Section~\ref{sec:solution} presents our automated query evaluation framework to
alleviate major issues in language ambiguity and help users find useful results more easily. Our
solution uniquely goes beyond existing code search recommendation approaches by
incorporating both static properties (query transformations) and dynamic runtime behavior (inspecting whether generated queries find results, which are then automatically displayed to the user).
\end{blockquoted}

\begin{blockquoted}
\textbf{Empirical results that demonstrate the effectiveness of our approach.}
In Section~\ref{sec:eval} we evaluate \approach with an A/B test across 10,000 unique users over 25 weekdays on our publicly-available
code search instance, \href{https://sourcegraph.com/search}{Sourcegraph.com}. We investigate click rate as a
positive indicator that users gain utility from code search and allows them to
make progress on their task at hand. Our main result shows that when \approach is active,
users are \textbf{22\%} more likely to click on result at all on any
given day, compared to the control group.
\end{blockquoted}

\noindent
We discuss related work in Section~\ref{sec:related} and conclude in Section~\ref{sec:conclude}.

\section{Building a Code Search Engine for All: Query Design Preliminaries from the Trenches}
\label{sec:background}

Sourcegraph aims to provide code search to the broadest spectrum of users, from
beginner programmers who have never written regular expressions, to power users
who can craft extremely sophisticated queries (e.g., search for a regular
expression in \ts{package.json} files only in repositories that have been
committed to in the last month).
At present, Sourcegraph interprets queries precisely with respect to a simple
language grammar. Broadly, a search query contains space-separated
\emph{patterns} (like \codeblue{func parse}) and \emph{filters} (like
\codepurple{path\codegray{:}\codeblue{package.json}}). Filters
specify a field prefix, like \codepurple{path\codegray{:}}, followed
by a value like \codeblue{package.json} to specify that only file
paths matching \codeblue{package.json} should be searched. Boolean
operators like \codepurple{AND}, \codepurple{OR},
\codepurple{NOT} apply to patterns and filters to build expressions.
This choice of syntax is largely consistent and conventional among major code
search tools, similarly found in Google Code Search, GitHub Code Search, and
OpenGrok. Although opportunities exist to design engines that interpret queries
more freely (e.g., as prompts to a machine-learned model), the state of code
search today has established the value of consistent and well-defined query languages to precisely find results and craft queries with extensive and predictable expressive power.\footnote{Albeit sometimes unforgiving to newcomers.}
We therefore scope our discussion to enabling greater ease of use when designing
such a well-defined language when talking about query interpretation. Within this scope,
we share our perspective on technical factors influencing code search usage with respect to query design.

\begin{table*}[t!]
\centering
\small
\caption{Query syntax properties in major code search engines today, and implications for language design and usability.}
\label{tab:1}
\begin{tabular}{p{40mm}p{131mm}}
\toprule
\bf Property          & \bf Design Implication                                                        \\
\rowcolor{LLGray}
single search string           & strive for terse expressive power and readability (e.g., choice of attributing semantic meaning to spaces in different contexts)  \\
support searching punctuation  & requires escaping or disambiguation (patterns conflict with reserved syntax in query language)  \\
\rowcolor{LLGray}
support sublanguages \newline (e.g., regular expressions) & requires escaping or disambiguation (dialect syntax conflicts with literal interpretation or query language)  \\
enable collaborative sharing & choice of consistent query syntax and semantics (it is exceedingly complex to support individual preferences for interpreting syntax or disambiguating; avoided by code search engines today) \\
\bottomrule
\end{tabular}
\end{table*}

\emph{What is special about query syntax for code search?}
Designing a code search language presents unique domain challenges in the
intersection of traditional search engines and full-fledged programming
languages. We draw on our experience and a cursory survey of related code
search tools~\cite{google-chromium-search}, \cite{gh-beta-search}, \cite{opengrok-lucene} to reveal common characteristics in this overlap.
Table~\ref{tab:1} distills these characteristics. In general \textbf{Properties}
add constraints to language design, which can lead to greater burden and
potential friction when users interact directly with a query syntax. Consequently all major
industrial code search engines today are influenced by, and account for, query
\textbf{Design Implications} in some shape or form. While not meant to be
exhaustive, these properties represent the most challenging aspects of language design in
our experience, because they constrain design that must be somehow reconciled with
user desires and expectation. They lie at the root of long internal discussions on how code search ought to work,
and give rise to much user confusion and misunderstandings (``Why do I see these
results? Why don't I see \emph{any} results? How do I set my preference to
interpret patterns as regular expressions by default?''). We elaborate on these properties below.

\emph{Singular search query strings and punctuation.}
Like Google search or other traditional search engines, code search
UIs generally offer a single input field (or maybe a small set of input in the
case of OpenGrok). In contrast, searching literally over code instead of web
content means that users are much more likely to search for literal punctuation
like quotes (as in our leading example), or other syntax that correspond to code
constructs: \codeblack{i < j, parse(, struct \{, !flag}. This is significant
because searching for punctuation increases the likelihood that search terms
conflict with reserved syntax in the query language (e.g., parentheses for
grouping search expressions like \codeblack{(foo OR bar)}). Further, and
unlike plain text search, major code search engines~\cite{google-chromium-search}, \cite{gh-beta-search}, \cite{opengrok-lucene}) effectively embed
support for sublanguages like regular expression syntax. Supporting such
dialects also heighten the potential for syntax conflicts with search terms, or
the query language itself.

\emph{Semantically and contextually significant spaces.}
Code search queries strive to be terse and less complicated than full-fledged
programming or database languages, but share overlap in expressive power
(boolean operations, conditional clauses). Most queries fit on a single line,
and whether by consequence or design, a majority of today's code search
engines apply some semantic meaning to spaces in the interest of
convenience and readability. The semantic meaning of spaces is a design choice, and may
differ across engines and between query terms. Some choices include spaces to
mean any of: (1) search for multiple terms anywhere in a document (without
respect to ordering); (2) search for terms \emph{with} respect to ordering
(e.g., on the same line); or (3) just lexically separate search filters.
Ultimately, shorthand conventions allow users to write more terse queries like
\setlength{\fboxsep}{1.5pt}\fbox{\small\texttt{repo:ase path:.tex table}} instead of
\fbox{\small\texttt{repo:ase {\color{purple}AND} path:.tex {\color{purple}AND} table}}. Such decisions
can generally enhance usability, but may sacrifice the precise intent of a user
in other contexts and introduce ambiguity (we elaborate in
Section~\ref{sec:challenge}).

\emph{Users want it their way, but that leads to bigger problems.}
The terse nature of search queries tend to invite ambiguity, and we've seen users
prefer different defaults for how queries should be interpreted. Drawing on our
previous example, users may prefer that quotes are interpreted literally, while
others may not. Instinct suggests that the search tool could provide a way for
users to configure personal defaults. In other words, the tool grants users
some agency over the query semantics. In practice we find this
idea does not work. Code search engines today are useful in collaborative
settings, where developers or coworkers share links to search results, and
queries are the source of input and truth. If the interpretation of that input
diverges across users, we violate a fundamental tenet: consistent, predictable
results. URLs or other serializable formats could in theory encode values that
indicate the intended (perhaps custom) semantics of a particular query to
overcome this challenge. In practice, no code search engine to our knowledge has
been able to identify a user-friendly way to ensure users copy a URL or share a
query format that encodes how query semantics. It is simply too easy and natural
for users to copy a query string in the search input box and share it as-is. In
this light, it is exceedingly difficult to cater to individual preferences for
query interpretation. Thus, in identifying mechanisms to reconcile ambiguity,
customizing query interpretation is not practical for us. Syntax needs to
be interpreted consistently to avoid further confusion, and reconciling
potential ambiguity requires a different approach.

\section{Challenges in Ambiguity: Observing the Pitfalls that Users Experience}
\label{sec:challenge}

\begin{table*}[t]
\centering
\small
\caption{A qualitative summary of how users experience pitfalls due to ambiguity in search query intent. The ambiguous columns are \emph{interchangeable}, and we've seen users express a preference for either interpretation. As a matter of consistent language design, a tool can typically implement only one sensible default, and sacrifices syntactic convenience for expressing alternatives..}
\begin{tabular}{p{20mm}p{36mm}p{50mm}p{55mm}}
\toprule
\bf           &          & \mc{2}{c}{\bf Ambiguity}                                                     \\
\bf Query     & \bf Pitfall & \bf Us: We guess you want to... & \bf User: I actually want to...            \\
\rowcolor{LLGray}
{\small\texttt{func parse}}
& Is term order significant?
& find matches of \texttt{func {\color{blue}AND} parse} anywhere in a file \newline
$\rightarrow$ ordering \emph{doesn't} matter
& find the string \texttt{func\textvisiblespace parse} on a single line \newline
$\rightarrow$ ordering matters; the pattern corresponds to a specific function definition \newline
\\
\texttt{{\color{blue}"}v1.3{\color{blue}"}}
& Are quotes meaningful?
& find \texttt{v1.3} \emph{without} quotes
& find \texttt{"v1.3"} \emph{with} quotes inside config files \newline
\\
\rowcolor{LLGray}
\texttt{swing{\color{blue}.*}}
& Interpret as regular expression dialect?
& match \texttt{swing} followed by anything \newline
$\rightarrow$ interpret \texttt{\color{blue}.*} as a wildcard regular expression.
& find wildcard imports in Java files \newline
$\rightarrow$ interpret \texttt{\color{blue}.*} literally, \emph{not} as a regular expression. \newline
\\
\texttt{func{\color{blue}.*(}}
& Interpret as regular expression dialect?
& try search for an regular expression, but it's \emph{invalid} \newline
$\rightarrow$ Tell user \texttt{\color{blue}(} might need to be escaped. Or did the user intend to search literally? \newline
& search a valid regular expression \texttt{func{\color{blue}.*\textbackslash(}} \newline
$\rightarrow$ User didn't realize need to escape \texttt{\color{blue}(}
\\
\rowcolor{LLGray}
\texttt{should {\color{blue}not} fail}
& Interpret \texttt{\color{blue}not} as a boolean operator or as a literal search term? 
& find matches in files for \texttt{should}, but exclude them if they contain \texttt{fail}
& find error messages and search literally for \texttt{should not fail} \\
\bottomrule
\end{tabular}
\label{tab:2}
\end{table*}

This section lays out practical examples and ``fork in the road'' decisions
that have accompanied the design of our industrial-strength code search query language
at Sourcegraph. We regularly receive user feedback from our customer
support channels and in-app feedback box. In the last three years we've received
hundreds of pieces of user feedback on experiences, opinions, and frustrations
specifically dealing with query syntax and semantics. We condense this feedback
qualitatively in Table~\ref{tab:2}, presenting the most common query \textbf{Pitfalls} users
experienced. Every pitfall leads to \textbf{Ambiguity} with more than one
plausible interpretation, and we give representative scenarios where a design
choice in our tool behaves contrary to a user expectation or desire. The
difference in our design choice (\textbf{Us} in the table) versus expected
behavior (by the \textbf{User}) is \emph{interchangeable}. In fact,
Sourcegraph currently supports two distinct modes for interpreting queries, due to
some of the contentious issues in ambiguity shown here.\footnote{Roughly, one
mode supports regular expression syntax as a first-class interpretation, and the
other prefers literal syntax. UI toggles allow switching between modes.} We elaborate on the concrete examples below.

The foremost behavioral pitfall in Sourcegraph is whether a sequence of
patterns like \fbox{\codeblack{func foo(bar, baz}} means ``search for each space-separated
term anywhere in a file'' (ordering does not matter) or whether the whitespace
should be interpreted literally on a single line (ordering matters). The same user may
even prefer one behavior over the other in different contexts, e.g., searching in a
more fuzzy way, versus pinpointing exact code. Language features exist to enable
both possibilities, but users may need to manually reformulate queries to achieve
what they want. For example, patterns can
be explicitly separated by \codeblue{AND} to express that ordering does not matter, or alternately, users can rely on the
default behavior that concatenates patterns, or users can quote patterns in a
filter \fbox{\codeblack{content:"func foo(bar, baz)"}} for absolute clarity and with
additional effort.

As in the leading example, users may start using Sourcegraph with a preconceived
idea of whether quotes are significant or not. Many have found it useful to
search quotes literally (cf. Table~\ref{tab:1}, supporting search for
punctuation) while others rely on quotes to express exact intent or to
disambiguate a regular expression. Because we want to maintain consistent
behavior (cf. Table~\ref{tab:1}, collaborative sharing), this is not a
preference we can accommodate in an individual user setting. Instead, users may
experience friction until they become familiar with the default behavior, and
mechanisms to change the behavior (e.g., UI toggles or additional syntax to disambiguate intent).

Sourcegraph currently supports the RE2 regular expression
dialect.\footnote{\href{https://github.com/google/re2/wiki/Syntax}{github.com/google/re2/wiki/Syntax}}
Supporting this sublanguage (cf. Table~\ref{tab:1}) leads to user confusion both
on (a) whether RE2 metasyntax is interpreted by default and (b) the realization that a
regular expression syntax may be invalid (and inconveniently so!). It is
generally difficult to infer whether a user intends their pattern to be
interpreted as a regular expression. We might speculate that patterns containing
syntax like \codeblue{\textbackslash w+} or \codeblue{.*?} is indicative,
but in practice the opportunities for ambiguity and conflict are too high to
\emph{override} the default behavior heuristically.

\begin{figure*}[t!]
 \centering
 \includegraphics[scale=0.412]{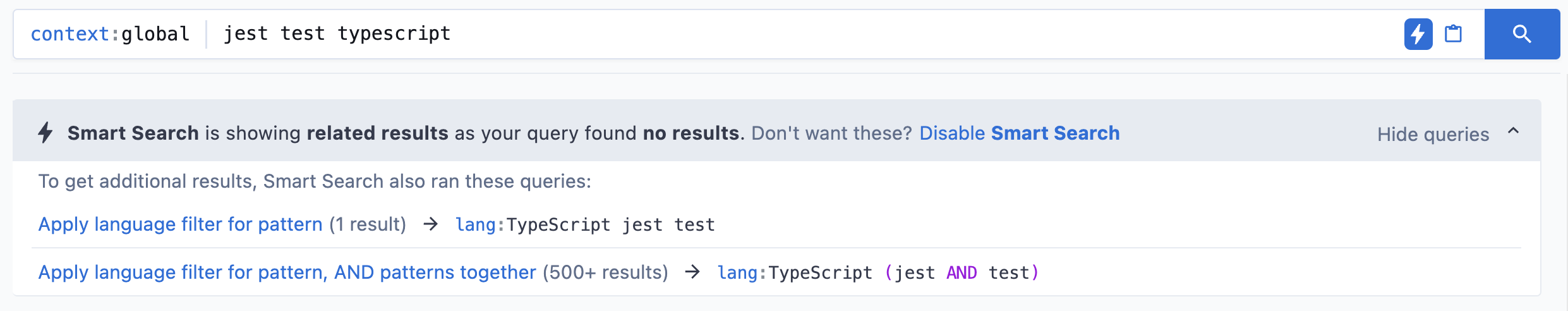}
 \caption{Search query input with a dialog box below. The \texttt{Smart Search} dialog box illustrates our search feature that implements \approach. This dialog box displays when a query like \setlength{\fboxsep}{1.5pt}\fbox{\small\texttt{jest test typescript}} does not return results \emph{and} we identify that it's possible to interpret the query in different ways based on our rules. The first rule detects that we can interpret {\small\texttt{typescript}} as a language. We display the corresponding action to {\codeactuallyblue{Apply language filter for pattern}} and the resulting query, which prepends a {\small\texttt{lang}} filter to form {\small\texttt{lang:TypeScript}}. Note that unlike traditional static query suggestions, our engine \emph{dynamically evaluates} alternative queries and \emph{only} reports the action taken if at least some results exist for an alternative. For example, we report the number of results (e.g., {\small\texttt{1 result}} associated with the first alternative, and {\small\texttt{500+ results}} for the second) in prioritized order. The results of these queries are also shown immediately below the dialog box (omitted in the figure for brevity) giving users a list of results without requiring them to click on suggestions.}
 \label{fig:smart-search}
\end{figure*}

A final balancing act involves adding more expressive power and syntax to the
query language. In most code search engines (Sourcegraph being no
exception) query strings are singular inputs (cf. Table~\ref{tab:1} single
search string). This grants some usability convenience, limiting the need for
interactive inputs like moving the cursor to multiple input fields or excessive
quoting in a command line. The decision carries the tradeoff that every addition
to the language, such as keywords or punctuation, increase the likelihood of
ambiguity with what a user may intend to search for literally. In practice,
visual cues alleviate ambiguity for cases like these (e.g., Sourcegraph
highlights keywords like \codepurple{NOT}). Unfortunately this still
requires users to reformulate their query when keywords or punctuation conflict
with the desired meaning.

\section{Mechanizing Auto Query Evaluation to Straddle Ambiguity}
\label{sec:solution}

Sections~\ref{sec:background} and \ref{sec:challenge} cover the challenges
we've faced to reconcile user expectation and tool design.
We realized that we cannot establish a set of default query behaviors that suit
all users. At best, we might converge on useful defaults for most users, and a substantial number of
remaining users will invariably experience some friction until they identify the
ways to disambiguate intent in our language (or, at worst, abandon the tooling).
These observations motivate us to consider a fresh perspective so that tool
behavior could straddle ambiguity and better accommodate user expectations. We
developed \fullapproach (\approach) which automatically runs queries under
alternative interpretations. This approach differs substantially from current
\emph{``Did you mean...''} suggestions found in typical code search tools today. Standard
practice in code search tools today implement checks that predominantly rely
purely on a query's \emph{syntactic} properties (e.g., check whether the pattern contains
quotes, or whether it matches a value like \codeblack{Python}, which might mean the user
only wants to search over Python files). However, because users often have
varying expectations of query \emph{semantics} (Section~\ref{sec:challenge}),
static checks can only partially alleviate user confusion or friction.
\fullapproach fills this gap by first requiring that alternative, suggested
queries actually produce one or more results before displaying this alternative
to the user. It then goes a step further by also \emph{showing those results
immediately} (up to some threshold of results) without requiring the user to
click on the suggestion or reformulate the query.

In contrast, simple query suggestions typically do not guarantee whether a
suggestion returns results over a data set (and so, if a user accepts the suggestion it may
return no results and not help them at all). Even then, users must respond to suggestions manually
(click on a suggestion or reformulate the query). With \approach, we develop a
framework to write rules that transform the user's original query \emph{and
evaluates alternatives at runtime} to ensure that suggested queries produce additional results that
the user may find useful. The approach seems straightforward and promising, but
it is deceptively difficult to implement well in an industrial-strength tool
that affects the day-to-day workflow of tens of thousands of developers. We next discuss our goals
behind \approach and its implementation challenges.

One overarching goal is to ensure \textbf{minimal disruption to existing
  behavior} while developing \approach. Users are sensitive to stark changes in
tools they use frequently, and our primary interest is to overcome disappointing
user experiences in the ``gray area'' of what a search query means while staying
true to any existing, learned expectations. We identified that the area of
greatest opportunity exists where user queries syntactically conform to a known
\textbf{Pitfall} (cf. Table~\ref{tab:2}) \emph{and} when a user sees \textbf{no
  results} for their original query. We can identify queries that experience
potential pitfalls \emph{statically}, but we must rely on \emph{runtime
behavior} to determine whether an alternative query may yield results. Our
implementation thus transforms the user's original query to try alternative
interpretations, but stipulates that the result of evaluating those queries must
return results before displaying a viable alternative and its outcome. Fig.~\ref{fig:smart-search}
shows the user-facing component that implements \approach.

\begin{table*}[t]
\caption{The main query transformation rules we evaluated with \approach. These mechanize the alternative interpretations associated with common pitfalls and frustrations we identified in user feedback (Table~\ref{tab:2}).}
\begin{tabular}{p{18mm}p{60mm}p{90mm}}
\toprule
\bf Rule name     & \bf Example application & \bf Description           \\
\rowcolor{LLGray}
\codeblack{and}
& \si{func\textvisiblespace parse} $\rightarrow$ \codeblack{func} \codeblue{AND} \codeblack{parse}
& converts ordered, space-separated patterns to an expression that searches for files containing all of those patterns in any order
\\
\si{unquote}
& \codeblack{"v1.3"} $\rightarrow$ \codeblack{v1.3}
& searches a quoted pattern without quotes
\\
\rowcolor{LLGray}
\si{regex}
& \si{func.*parse} $\rightarrow$ \si{/func.*parse/}
& heuristically interprets a pattern as a regular expression, rather than literally.\footnotemark
\\
\si{language}
& \si{python} $\rightarrow$ \si{language:python}
& converts a pattern to a filter that restricts search to files of a particular language
\\
\bottomrule
\end{tabular}
\label{tab:rule-breakdown}
\end{table*}
\footnotetext{Roughly, we detect whether two or more metasyntax operators exist. The full implementation is \href{https://sourcegraph.com/github.com/sourcegraph/sourcegraph@21c49c855ea9fc0deb9bc32e09d6a0e14649d500/-/blob/internal/search/lucky/rules.go?L107-174}{available online}.}

\subsection{Building a lazy query generator}

A subsequent goal of \approach is to accommodate a \textbf{configurable suite of
  potential transformation rules} that can overcome query pitfalls. We
identified a recurring set of issues based on user feedback (Table~\ref{tab:2})
and seek a way to encode multiple rules and to test their effectiveness. We've
mechanized \approach with a set of atomic rules summarized in
Table~\ref{tab:rule-breakdown} that correspond to the most problematic pitfalls.
Rules are implemented in Sourcegraph as query passes using a visitor framework,
written in Go. It is natural for multiple atomic rules to apply independently to
a user's original query, and it is also natural for atomic rules to compose. The
second suggestion in Fig.~\ref{fig:smart-search} illustrates an instance of
composite rules: apply a \codeblack{language filter} to \codeblack{typescript}
and then also convert the remaining patterns to an \codeblack{AND} expression.
\approach achieves this power by feeding atomic rules to a lazy query generator.
The generator applies rule transformations in \emph{prioritized order} and
iteratively attempts combinations of atomic rules in the same order. The order
of generated queries bear on how alternative searches are evaluated at runtime,
and we explain why this is significant shortly. To curb combinatorial explosion
during query generation, we first prune rules that cannot apply to a query
(where rule preconditions are unsatisfied) and bound the number of valid query
alternatives to evaluate (experimentally, a threshold of 5 queries work
well in practice).

Thus far, we've described rules implemented in \approach. Rules specify (only)
static preconditions and transformations for generating alternative queries. A
lazy generator encodes the sequence of statically valid \emph{candidate} queries
that may produce results that are helpful to the user. Merely suggesting the
possible alternatives in the traditional sense may overwhelm users (if the
generator produces many candidates) or produce unhelpful ``no op'' alternatives
if those queries do not return results once they are evaluated. This is where
\approach takes a step further by iteratively evaluating candidate queries,
inspecting whether there are any results, and returns these to the user.

\subsection{Implementing and taming runtime behavior}

A crucial practical goal is to additionally ensure \textbf{performant runtime
  behavior} when evaluating alternative queries. Our backend search service is
built in key ways to enable \approach to work well at runtime. First, results
found ordinarily by our search engine are immediately \emph{streamed} to the client. This
has the benefit that users see results before the entire search query
necessarily completes (users experience a quicker time-to-first-result than waiting for the search to complete). When evaluating
alternative queries, supplementary results are streamed only \emph{after} evaluating
the original query. Our backend counts the number of results streamed by the
original query, and then decides to run alternative queries if the original
query returned either \textbf{no} results or if the original query returned
\textbf{some} results, but fewer results than we are willing to display to the
user. The threshold of results we show to the user is 500. Thus, if the original
query returns 500 or more results, we do not run \approach. If less, we run
\approach, which appends supplementary results corresponding to alternative
queries (if any) until we reach the threshold of 500 results or exhaust the
number of alternative queries to run. This behavior ensures minimal disruption
with respect to the original query's meaning, and only reports additional
results under the special condition where alternative query results supplement
the original behavior. Equally critical, we immediately shortcircuit running \approach
once the maximum threshold of results are streamed, circumventing performance
issues due to long-running queries (and awaiting their
outcome) or large in-memory result sets.

We evaluate queries sequentially in the order they are generated lazily. As a
further step we've also decoupled query generation and evaluation such that our
search architecture can run multiple alternative queries in parallel---a
characteristic that helps us explore tradeoffs in performance and precision for
supplementary results. We implement the evaluation order of atomic rules based
on our intuition of how those rules may grow the supplementary result set.
Roughly, we prefer rules that are \emph{more particular} (or specific) to fire
first, followed by more general rules that require looser conditions to
apply. Table~\ref{tab:rule-breakdown} shows the rule order in reverse, from most
general (\codeblack{and}) to less general (\codeblack{language}). The intuition
is that applying more particular rules, e.g., \codeblack{language}, are likely to narrow results and produce a relatively smaller result set compared to other rules, since it has a restrictive effect on the overall search space. On the other hand, more
general rules like \codeblack{and} remove an ordering constraint on queries and cast a wider net over the search space. Such looser constraints tend to produce relatively larger result sets and can easily trigger the 500+ result threshold.
Intuitively, if \codeblack{and} rules are prioritized and produce a flush 500+
result set, we may never attempt more specific rules like \codeblack{language} and miss the opportunity assist users during more particular query pitfalls.

While there are clear future opportunities to consider additional static rules
or runtime behaviors, our initial feature set entailed the implementation described. To measure its effectiveness we ran an A/B search on our public code search instance.

\section{Evaluation}
\label{sec:eval}

\begin{figure*}[t!]
\begin{subfigure}{.9\columnwidth}
\centering
\includegraphics[scale=0.6]{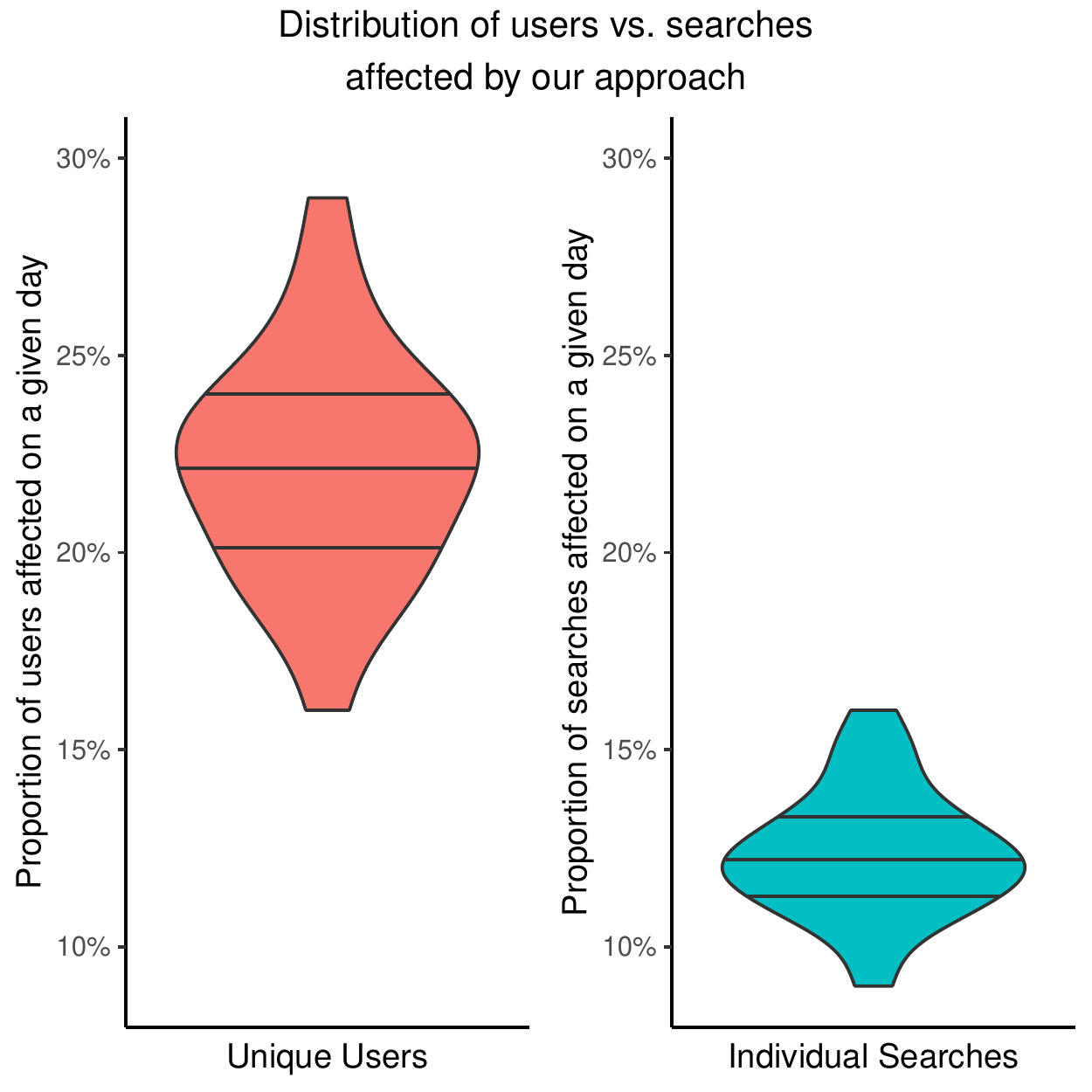}
\caption{The effect of activating \approach. On average 22\% of unique users activates \approach with one or more of their searches (left). In terms of individual searches, \approach activates 12\% of the time on average. At most 29\% of users and 16\% of searches are affected respectively, which helps forecast the magnitude of potential benefit or degradation in experience we can expect to impact with \approach.}
\label{fig:2-effect-distro}
\end{subfigure}\hspace*{4em}
\begin{subfigure}{.9\columnwidth}
\centering
\includegraphics[scale=0.6]{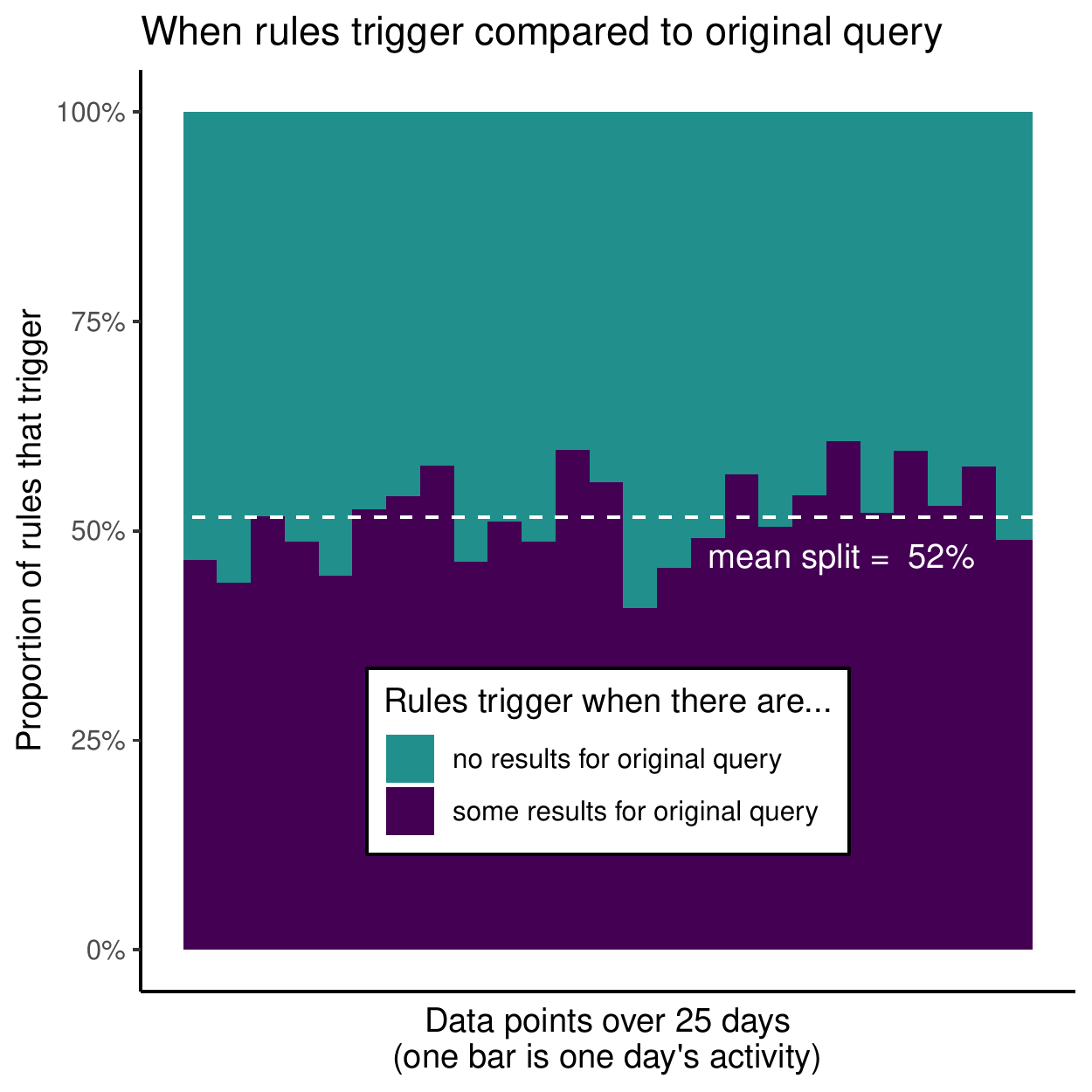}
\caption{How \approach behaves compared to a user's original query. On average \approach triggers 52\% of the time when the original query had \textbf{some} existing results, and 48\% of the time when the original query had \emph{no} results. Knowing the proportion of searches that yield no results is an especially appealing area to improve user experience, and 48\% provides ample opportunity for \approach to activate.}
\label{fig:2-trigger-distro}
\end{subfigure}
\caption{}
\label{fig:2-composite}
\end{figure*}

We evaluate our approach with an A/B test conducted on
\href{https://sourcegraph.com/search}{Sourcegraph.com}, our publicly-available
code search instance. This instance indexes over 2 million of the most common
open-source repositories hosted on GitHub and GitLab. Our overall objective is
to evaluate whether automatically evaluating alternative queries help users
find results they are interested in. We use \emph{click rate} as a standard measure to gauge
whether users find results they are interested in. In a typical user flow,
users enter a search query and then see a list of results. They may
subsequently click on a result in the list, which redirects the user to an
expanded file view positionally anchored at the clicked result. We make the
underlying assumption that when a user clicks on a result, it is relevant and
useful to their task at hand.\footnote{See, e.g.,~\cite{howcodesearch} for example tasks where
software developers use code search.}

We can only observe a meaningful delta in click rate if \fullapproach (\approach) actually
triggers for the user's query. Two conditions
must be met for \approach to trigger: (1) one or more rules must apply to the
original query and (2) the original query returns fewer results than we are willing to display
(i.e., there is room in the web client to display more results that may be useful).
Thus, as a first step to evaluating click rate, we are
initially interested in whether the number of queries where \approach fires is
significant at all, considering our conservative choice of triggering conditions.
Our research questions are as follows:

\nbf{RQ. 1: Do a significant proportion of user queries trigger \approach?} If
we can reject the null hypothesis that \approach has no observable effect, then
we're also interested in the proportion of queries that trigger \approach. The proportion
of queries that trigger \approach establish to what extent we may impact a user's
search experience. Consequently, we also ask:

\nbf{RQ. 2: Does user click rate differ between \approach users and the control
  group?} If we can reject the null hypothesis that \approach has no observable
effect on click rate, then we're also interested in (a) the magnitude of the
delta in click rate and (b) the distribution of \approach rules corresponding
to clicked results.

These research questions inform strategies to improve a user's search experience
and the utility they gain from it. For example, \textbf{RQ. 1} indicates not only
to what extent we might achieve a positive outcome, but also flags the
possibility that \approach may apply too eagerly and produce unpredictable
effects (rules are ``noisy'' and produce noisy results). Similarly, \textbf{RQ.~2} indicates the behavior and efficacy of various rules (if any), informs
design tradeoffs to consider (apply conservative versus aggressive rules in
different contexts).

\subsection{Experimental Setup}

Our A/B test ran for 25 weekdays (5 weeks, consistent with the range of standard
practice~\cite{KohaviLSH09}) across more than 10,000 unique users, averaging
approximately 1,600 unique users per day.
We activated the new automated query suggestion feature for half of the population
($\approx$800 unique users per day), with the remaining 50\% of users
establishing the control group. Feature activation is consistent and
deterministic per user, meaning that once a user is bucketed into either the A
or B variant, they receive the same feature set over the duration of the A/B
test. Users who receive the new feature variant are exposed to the search
behavior and dialog box shown in Fig.~\ref{fig:smart-search}.

We collected anonymized, aggregate data over the course of the A/B test. To
answer our RQs we instrumented the application to collect whether a
user's query triggered \approach, which rules applied to their query, and
whether they clicked on a result associated with alternative queries. Each
search event is further associated with one of the two categories:
(1) whether the event occurred when the user's original
query returned \textbf{no} results, or conversely (2) whether the event occurred
when the user's original query did return \textbf{some} results, but fewer than
the maximum number of results we are willing to display. We consider these
distinct scenarios to better understand the hypothetical utility of rules and
result clicks. In the first category, we know that a clicked result must
correspond to results generated \emph{purely} from \approach. In the second, we
know that a result click corresponds to a result set to which \approach
\emph{added} results, but due to the instrumentation complexity we do not record
whether the clicked result corresponds to the original query or an automatically
generated query (it may be either).


\subsection{Experimental Results}

\nbf{Effect on user search experience (RQ. 1).} Our results show that an average of \textbf{22\%} of unique users in the experiment group trigger \approach per day. That is, on a per-user basis, at least one of the user's searches triggers \approach. In terms of all search events, an average of \textbf{12\%} of individual searches trigger \approach. Fig.~\ref{fig:2-effect-distro} summarizes this result. For both observations we assume normality (Shapiro-Wilk test, $p > 0.5$) and conclude significance (one sample t-test, $p < 0.5$).
We observe 4.19 searches per user in the control group versus 3.95 searches per user with \approach.
In terms of the \emph{number of searches} per user, we forego a deeper analysis to meaningfully conclude whether \approach is significant: The fractional difference of 0.24 searches per user has little impact on our \text{RQ}s, and we suspect there exists significant variance in usage here
that would require additional instrumentation to control for (e.g., heavy users versus novice users, or even automated scripts).
%
%
%
However, taking the overall information as a rough indicator, observing $\approx$4 searches per user suggests that in the distribution of users compared to searches (Fig.~\ref{fig:2-effect-distro}), \approach has a significant probability of triggering per user,
and affects one or more of a typically small number of searches.

Fig.~\ref{fig:2-trigger-distro} summarizes our subsequent analysis to bucket how
\approach behaves compared a user's original query. Whenever a user triggers
\approach we record whether their original query returned \textbf{no} results or
\textbf{some} results. On average \approach triggers slightly more often when the
original query produces some results (\textbf{52\%}) versus no results
(\textbf{48\%}). Intuitively, \approach provides greater utility when the
original query produces no results, since we then create
opportunity for users to see or click on results that they
would not have had otherwise. Our key take away is that the 48\% proportion of searches in this category suggest ample opportunity to leverage \approach in the worst ``no results'' case.

\begin{figure*}[t!]
 \centering
 \includegraphics[scale=0.68]{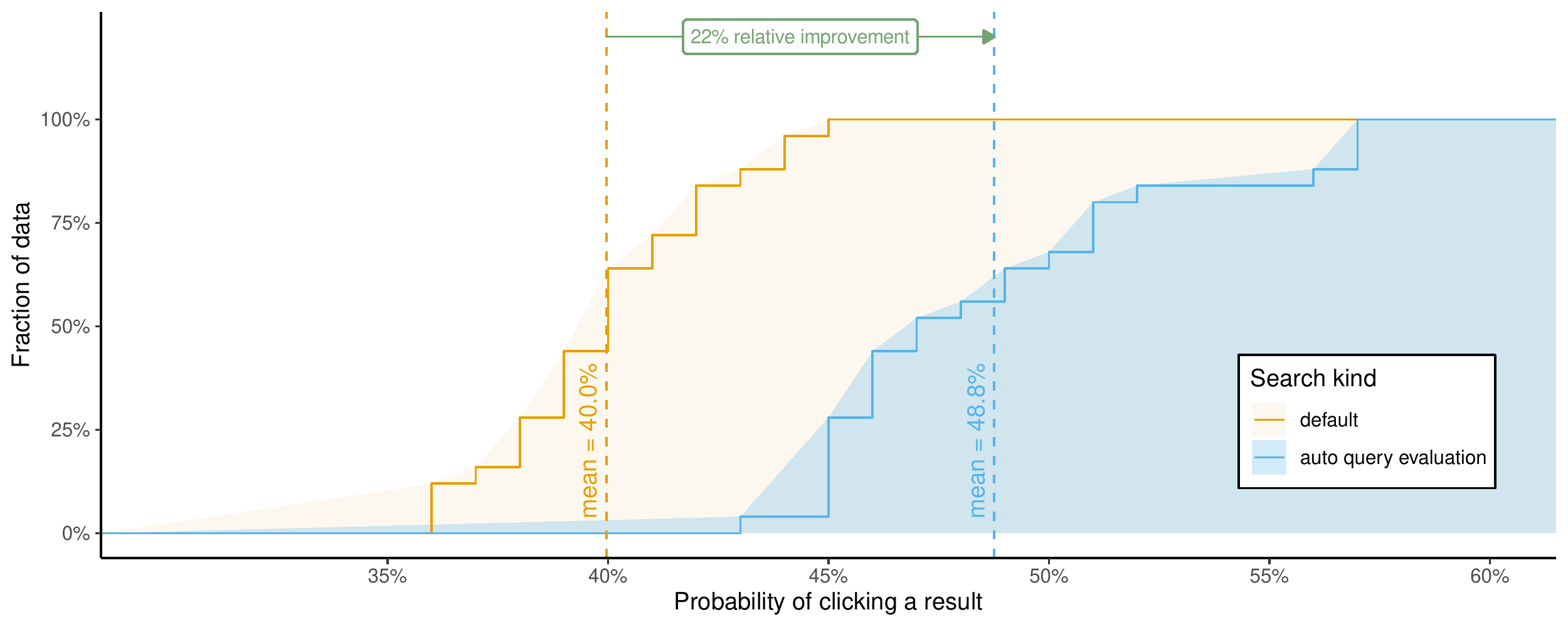}
 \caption{Cumulative distribution function over users who click on a search
   result at all on a given day. Our main result shows that on average, users
   are \textbf{22\%} more likely to click on a search result when \approach is active,
   relative to the baseline default behavior. There is also a notable uniform
   shift across the distribution, and users with \approach are more likely to
   click on a result \emph{on average} than the
   \emph{best-case scenario} for users in the baseline (compare the better 48.8\% mean for \approach to  the maximum probability of $\approx$45\% representing all users in the baseline group).}.
 \label{fig:1-main-result}
\end{figure*}

\nbf{Effect on click rate (RQ. 2).} The most meaningful aspect of our experiment
investigates whether our approach is ultimately helpful, which we pose by
asking: How does \approach affect click rate? We observe the most compelling
result when we consider searches by unique user, and whether unique users click
any result at all on a given day.\footnote{Note that this data is a function of our
aggregation telemetry: we do not associate search events with each user, we simply count
whether they clicked a result at all (not how many times they clicked a result).}
Fig.~\ref{fig:1-main-result} summarizes our main result. On average, on any
given day, we find that users are \textbf{22\%} more likely to click on a result
at all compared to the baseline when \approach is active.
Interestingly, Fig.~\ref{fig:1-main-result} shows an almost uniform shift in the
distribution of clicked results by user when \approach is in effect.
 The shift is substantial:
users with \approach are more likely to click on a result
\emph{on average} than even the \emph{best-case}, highest-probability scenario for users without it (Fig.~\ref{fig:1-main-result}).

When we consider all search events in aggregate (without respect to unique
users, since our data does not associate the number of searches to
individual users), the effect of \approach is more subtle, but still notable. In
absolute terms, we find that the total number of clicked results relative to
searches decreases slightly by 2\% (two-sample t-test, $p \approx$ 0.04), but we must be
mindful that \approach operates by strictly adding results to what the original
query produces, and especially when the original query produces no results (cf.
Fig.~\ref{fig:2-trigger-distro}).

In other words, if users click on search results corresponding to queries that
would normally return \textbf{no} results, but they click on such results with
\emph{lower frequency} than the average click rate of queries that do return
results, we'll observe an overall drop in click rate, even when we are in the
desirable case where users are presented with more opportunities to click results they
otherwise would not have.
We found this behavior indeed influences user
clicks with \approach. For one, searches conducted with \approach active
yield non-empty results \textbf{10\%} more of the time compared to the baseline (which
correspondingly yields no results). Further, \textbf{5\%} of all clicked
results with \approach correspond to queries that ordinarily have \textbf{no}
results. In absolute terms, all clicks in this group are additive to the baseline click rate (\textbf{+5\%}) in practice,
because results correspond to searches that are ordinarily
``unclickable''. However, the click rate of \approach in the ``no result'' bucket is a lower
\textbf{33.3\%} compared to the baseline click rate at \textbf{49.3\%},
accounting for an overall relative drop in click rate. Similarly clicked results
corresponding to queries with \textbf{some} results account for \textbf{7.6\%} of
all clicks, with a click rate of \textbf{47.31\%}. We found it somewhat
surprising to see a slight drop in click rate for the case where \approach adds
results. Here we considered how the breakdown of rules corresponding to clicks might deliver
additional insight.

\begin{figure*}[t!]
 \centering
 \includegraphics[scale=0.72]{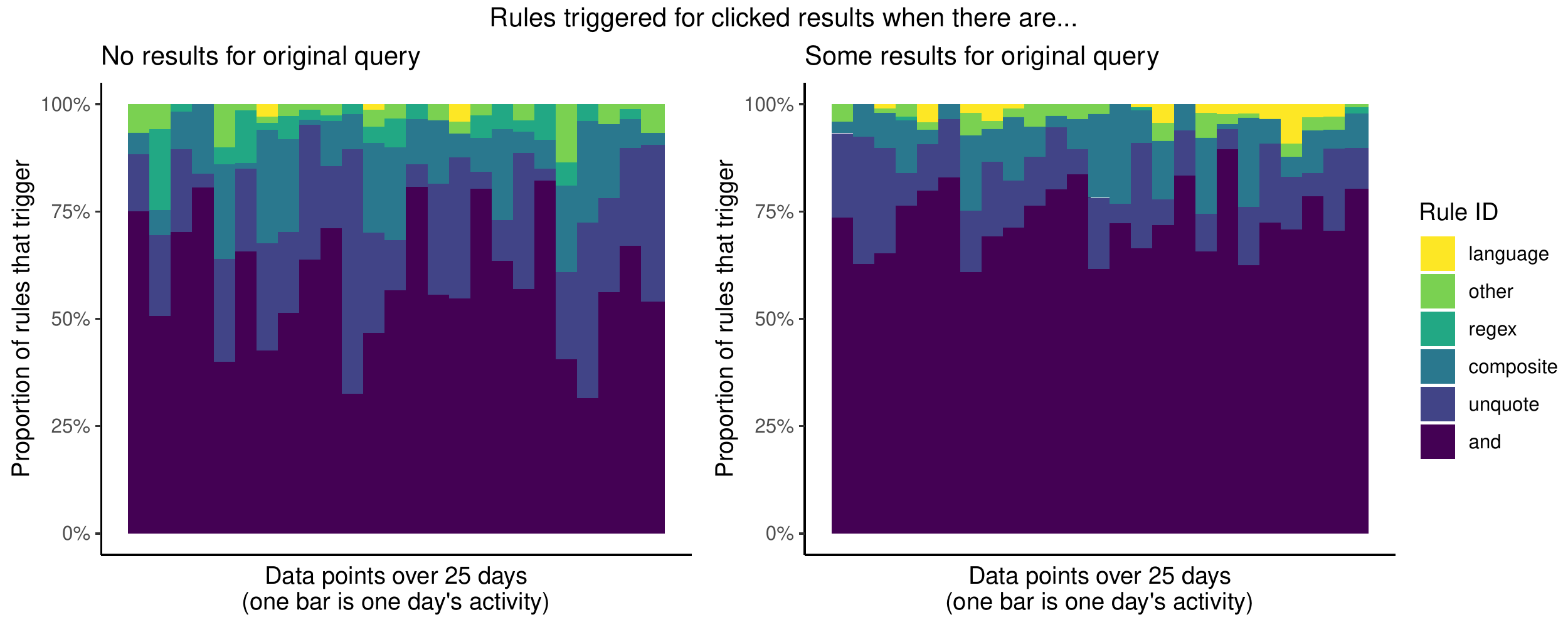}
\caption{Rule breakdown for clicked results. The \codeblack{and} rule dominantly triggers for queries that ordinarily have \textbf{no} results (left) and queries that do have \textbf{some} results (right). However, there is a stark difference in click distribution between these two groups, with the ``no results'' case (left) seeing a more varied tendency to click on results across rules. The difference may be due to inherent properties of the search patterns and files being search, user profiles (new users versus familiar users), or resulting from search usage in different contexts for varying tasks.
}
\label{fig:3-rule-breakdown}
\end{figure*}

Fig.~\ref{fig:3-rule-breakdown} summarizes which rules triggered a given query,
broken down by whether that rule was associated with an original query that had
\textbf{no} results (Fig.~\ref{fig:3-rule-breakdown}, left) versus \textbf{some} results (Fig.~\ref{fig:3-rule-breakdown}, right). Rule IDs correspond to the
transformation rules in Table~\ref{tab:rule-breakdown}. In addition we show rule
\si{other}, a catch-all identifier other experimental rules that we anticipated
would trigger less frequently (e.g., detecting whether a user wants to search
commit messages versus function symbols, or converting GitHub URLs to repository
filters). Rule \codeblack{composite} additionally represents any composition where two or
more of the rules apply
(e.g., apply the \codeblack{lang} rule followed by the \codeblack{regex} rule).
Due to the large number of possibilities, we do not currently record each
discrete combination of rules for composite queries (in future using an
efficient encoding may enable us to do so).

Rule \codeblack{and} triggers most often in both scenarios, where the original query returns either no results or some results. This is somewhat expected, since intuitively rule \codeblack{and} requires looser properties to trigger compared to other rules: a search query needs to contain just two or more search terms like \si{func parse}, and the search must find results where files contain both terms. Contrast this with rules that require more specific conditions to trigger
patterns (e.g., recognizing a language like \codeblack{Python} for rule \codeblack{lang}, or patterns containing quotes). However, we were surprised by the stark difference when comparing the distributions in Fig.~\ref{fig:3-rule-breakdown}. The proportion of clicks for the \codeblack{and} is lower for the the \textbf{no} results case on average, and the difference in variance is significant (F-test, $p < 0.05$).
Our data reveals that when the original query yields no results, there is a greater variation in rules that apply and corresponding results clicked (Fig.~\ref{fig:3-rule-breakdown}, left).
One possibility is that the terms in the query are inherently less likely to be found together in files for these kinds of searches (i.e., the \codeblack{and} rule succeeds less often with respect to code being searched). Conversely, cases where rule \codeblack{and} succeeds along with the original query (Fig.~\ref{fig:3-rule-breakdown}, right) may mean the search terms overlap and share the same search space, finding multiple satisfying results. Another possibility is that this difference accounts for different kinds of users. For example, it may happen that new users who are unfamiliar with the query language click results produced by the \codeblack{unquote} rule more frequently in the ``no results'' case, if they expect quotes to work similar to that of Google search. Segmenting these concerns is part of future work. Our main takeaway from the current data is that it is evidently worth evaluating the effectiveness of rules in these different contexts, and strategically picking the ones that are most useful contextually. As mentioned, we observed a slight drop in click rate for \approach in the case where there already are existing results. Fig~\ref{fig:3-rule-breakdown} concretely points out that the rule most likely responsible for this change is an aggressive application of the \codeblack{and} rule. This kind of information is key for testing rules and converging on a set of rules with a greater utility.

\subsection{Discussion and Limitations}

We were surprised by the substantial shift and impact in search and user behavior induced by \approach, considering our focus on minimizing disruption to existing behavior. One of our takeaways is that ``gray area'' intent with respect queries has a signficant effect on failure-prone behavior that end up yielding \textbf{no} results. We set up our experiment to segment and control for organic usage on weekdays~\cite{KohaviLSH09}, and rely on session cookies to identify unique users. The validity of our results could be affected by differing usage patterns between enterprise customers and our public, open-source instance. Additionally, usage patterns may differ between users of varying expertise. Segmenting users based on usage patterns is an area of future consideration. Implementing \approach performantly in other code search tools is an important consideration for product viability. Our evaluation shows that \approach triggers additional searches 10\% to 15\% of the time (inducing additional performance overhead) with the potential to assist an outsized number of individual users (up to 29\%). While attractive in our setting, this tradeoff will depend on tool context and deployment costs.

\section{Related Work}
\label{sec:related}

The most closely related work to ours informs how developers use code search at
Google~\cite{howcodesearch}. This case study covers how users interact with
filters like \codeblack{lang:}, and the kinds of software tasks that influence
how users create and reformulate queries. Similar to our investigation, this
work likewise suggests that search usage and clicks depend on varying developer
contexts and query properties. In contrast, our approach delves specifically
into altering the default search behavior, testing whether we can assist users
in finding results by automatically attempting alternative query
interpretations. In this light, \approach presents a new way to automatically
reformulate queries based on known pitfalls and complexities introduced by query
language design.

A recent survey covers a plethora of opportunities and challenges in code search
tools~\cite{codesearchsurvey}. Studies consider approaches to query
reformulation~\cite{HaiducBMOLM13,SismanK13} and techniques that influence
ranking~\cite{StoleeED16}. A large body of work investigates query expansion
to process natural language queries to find relevant code
snippets~\cite{HuangYC19}, \cite{LiuKMC019}, \cite{SirresBKLKKT18}. In general, while academic pursuits and studies offer promising future
directions for developer tooling, they do not operate with the constraint where tens of thousands of
active users may be drastically or negatively impacted by new approaches, nor do
they typically account for deploying a workable solution at our scale. We
developed our approach in direct response to vocal feedback where users
experience recurring papercuts in an already popular industrial code search
tool. We found that many query papercuts stem from interactions influenced by
language properties and confusion around intent and semantics. Our focus was on
reconciling this language confusion with \approach, rather than a focus on
increasing result relevance or ranking, or, e.g., building a natural language
query interpreter. In analyzing qualitative user feedback and consequently
developing \approach, we found that targeted query transformations can have a
significant effect on usability and search behavior while minimizing disruption
to our users.

\section{Conclusion}
\label{sec:conclude}

We identified unique challenges in designing a code search language and revealed
how design choices in query behavior can lead to mismatches in user expectation.
We developed \approach to help reconcile contradictory user expectations by
automatically evaluating query alternatives and showing those results to users.
We evaluated our approach over a large A/B test averaging 1,600 active users a day. We found that with \approach, users are on average 22\% more likely to click on a result at
all on a given weekday compared to ordinary users. Considering our focus on minimizing disruption to existing behavior, we were surprised by the substantial shift and impact in search and user behavior induced by \approach. Evidently, failure-prone cases that manifest due query language properties and apparent ambiguities can impact users extensively. We expect that investigating greater application of \approach can fruitfully assist users in avoiding pitfalls, and that its development can benefit today's code search tools at large.

\section{Acknowledgments}
\label{acks}

The author would like to acknowledge the help and support of Juliana Pe\~{n}a,
Paulo Almeida, and the Search Product team at Sourcegraph generally in the
design and implementation of the Smart Search feature related to this research.

\balance
\bibliographystyle{IEEEtran}
\bibliography{misc}

\end{document}